# E-Commerce and E-Health Strategies and Implementation Activities in the United Kingdom: Review Study

Solomiia Fedushko[1][0000-0001-7548-5856], Liubov Shevchuk [2][0000-0001-8057-6357],

Anastasiia Poritska[1], Ruslan Kravets[1][0000-0003-2837-9190],

Oksana Tymovchak-Maksymets[1][0000-0002-6044-3407]

[1]Lviv Polytechnic National University, Lviv, Ukraine
[2] Lviv University of Business and Law, Lviv, Ukraine

`solomiia.s.fedushko@lpnu.ua, ltshevchuk@ua.fm,`
`ahorychka@gmail.com`

**Abstract.** This study deals with the review of e-commerce and e-health. The investigation of e-commerce and e-health in the United Kingdom is proposed. The related works of e-commerce and e-health sectors is analysed. The world market for e-commerce is studied.

## 1 Introduction

E-commerce is the field of digital economy that includes all financial and trade transactions that are conducted through computer networks, and business processes related to these transactions. Initially, e-commerce relied on non-Internet communications and was characterized by the use of various stand-alone standards and protocols. The development of the Internet has led to a significant reduction in the cost of using e-commerce due to low cost of information exchange. It also stimulated the further development of e-commerce, among which today there are the following levels of its purpose: B2B - Business-to-Business is the whole electronic commerce of companies, B2C - Business-to-Customer is trade by private and private audiences or by business people and etc. In many cases, e-commerce reduces the resale of the product from producer to consumer. This is possible through the use of Internet technologies that enable effective direct interaction with the end user, so companies can play the role traditionally played by intermediate suppliers. It also allows you to accumulate information about all sales and all customers, which in turn allows you to perform a thorough business analysis and marketing research. This is a great competitive advantage. The biggest advantage of e-commerce in United Kingdom is a significant reduction in the cost of the transaction and its subsequent maintenance. Therefore, business processes that can be transferred to an electronic basis have the potential to reduce costs, which in turn reduces the cost of goods or services. The most famous example of e-commerce is an online store, which is a web resource with products catalog and the ability to order and buy products that likes buyer. More and more companies around the world are implementing e-commerce results in business. E-commerce includes electronic information exchange E-Banking, E-Cash, Electronic Funds Transfer, E-Trade, Electronis Data Interchange, E-Insurance, E-Marketing, E-health, etc. E-health is an important component of modern e-society and significant sections of e-commerce.

## 2  Related works

Researchers and specialists in the field of e-commerce pay a lot of attention to the study of this issue. An analysis of existing research papers on the implementation of e-commerce in the United Kingdom, which is presented in the table.

**Table 1.** Analysis of scientic works in e-commerce

| Scientific fields | Related works |
|---|---|
| e-commerce | — digital business and e-commerce management (Chaffey, Dave, Tanya Hemphill, David Edmundson-Bird)<br>— e-commerce: business, technology, society (Laudon, Kenneth C., Carol Guercio Traver)<br>— B2C e-commerce: impact in urban areas (Esser K., Kurte J.)<br>— The social and cognitive impacts of e-commerce on modern organizations (Khosrowpour, Mehdi)<br>— introduction to e-commerce (Rayport, Jeffrey F., Bernard J. Jaworski)<br>— recommender systems in e-commerce (Schafer B., J. Konstan, J. Riedl)<br>— e-commerce recommendation applications (Schafer B., Joseph A. Konstan R.)<br>— e-business and E-commerce Management: Strategy, Implementation and Practice (Chaffey, Dave) |
| e-commerce in the United Kingdom | — enforced standards versus evolution by general acceptance: A comparative study of e-commerce privacy disclosure and practice in the United States and the United Kingdom (Jamal K., Maier M., S. Sunder)<br>— digital signature law of the United Nations, European Union, United Kingdom and United States: Promotion of growth in E-commerce with enhanced security (Blythe, Stephen E.)<br>— digital divide and the quality of electronic service delivery in local government in the United Kingdom (Kuk, George)<br>— adoption of ICT in agricultural management in the United Kingdom: the intra-rural digital divide (Warren, Martyn F.)<br>— challenges and utilization of e-commerce: use of internet by small to medium-sized enterprises in the United Kingdom (Lawrence J. Eke) |

Also the detailed analysis of works in one of the main sectors of e-commerce – e-health in the world and the UK in particular, is conducted.

**Table 2.** Analysis of scientic works e-health fields

| Scientific fields | Related works |
|---|---|
| e-health | — e-health implementation toolkit: qualitative evaluation across four European countries (MacFarlane, Anne, et al.)<br>— barriers to e-health business processes (Mieczkowska, Suzanne, Matthew Hinton, David Barnes)<br>— M2M communications for E-health and smart grid: An industry and standard perspective (Fan, Zhong, Russell J. Haines, Parag Kulkarni)<br>— no more dithering on e-health: let's keep patients safe instead (McGrail, Kimberlyn, Michael Law, and Paul C. Hébert) |
| e-health in the United Kingdom | — implications of e-health system delivery strategies for integrated healthcare: lessons from England (Eason, Ken, and Patrick Waterson)<br>— E-health data to support and enhance randomised controlled trials in the United Kingdom (Harron, Katie, Carrol Gamble, Ruth Gilbert)<br>— E-health and Universitas 21 organization: Telemedicine and underserved populations (Wootton R., Laurel Jebamani, Shannon Dow) |
| mobile e-health | — mobile e-health: evolution of telemedicine (Tachakra S., et al.)<br>— mobile e-Health monitoring: an agent-based approach (Chan, Victor, Pradeep Ray, and Nandan Parameswaran)<br>— mobile e-health: Making the case (Archer, Norm)<br>— mobile e-health system based on workflow automation tools (Pappas, Christodulos, et al.)<br>— intelligent cloud-based data processing broker for mobile e-health multimedia applications (Peddi, Sri Vijay Bharat, et al.)<br>— evolution of E-Health–mobile technology, mHealth (P. Chandrashan)<br>— patient education on mobile devices: an e-health intervention for low health literate audiences (Mackert M., Brad Love, Pamela Whitten)<br>— new secure communication protocols for mobile e-health system (Aramudhan, M., and K. Mohan) |
| telemedicine | — telemedicine and e-health (Pinciroli, Francesco, et al.)<br>— e-health, telehealth, and telemedicine (Maheu Marlene, Ace Allen)<br>— from telehealth to e-health: virtual reality in health care (Riva G.)<br>— visualization of e-Health research topics and current trends using social network analysis (Son, Youn-Jung, et al.)<br>— telemedicine (Wootton, Richard)<br>— telemedicine technology, clinical applications (Perednia D., Ace A.)<br>— benefits and drawbacks of telemedicine (Hjelm, N. M.)<br>— essentials of telemedicine, telecare (Akhlaghi Hamed, Hamed Asadi)<br>— effectiveness of telemedicine (Ekeland A., Signe Flottorp)<br>— Telemedicine system (Tarassenko, Lionel, et al.)<br>— introduction to practice of telemedicine (Craig John, Victor Petterson)<br>— definition and evaluation of telemedicine (Bashshur, Rashid L.)<br>— wireless telemedicine systems (Pattichis, Constantinos S., et al.) |

The scientist Gasser U. et al. [26] proposed the sunburst diagram (Figure1) that shows 6 principles raise ethical and legal issues when considered in relation to digital public health technologies against COVID-19. As presented in the circles at the centre, these principles apply equally to symptom checkers, flow modelling, quarantine compliance, and proximity and contact tracing.

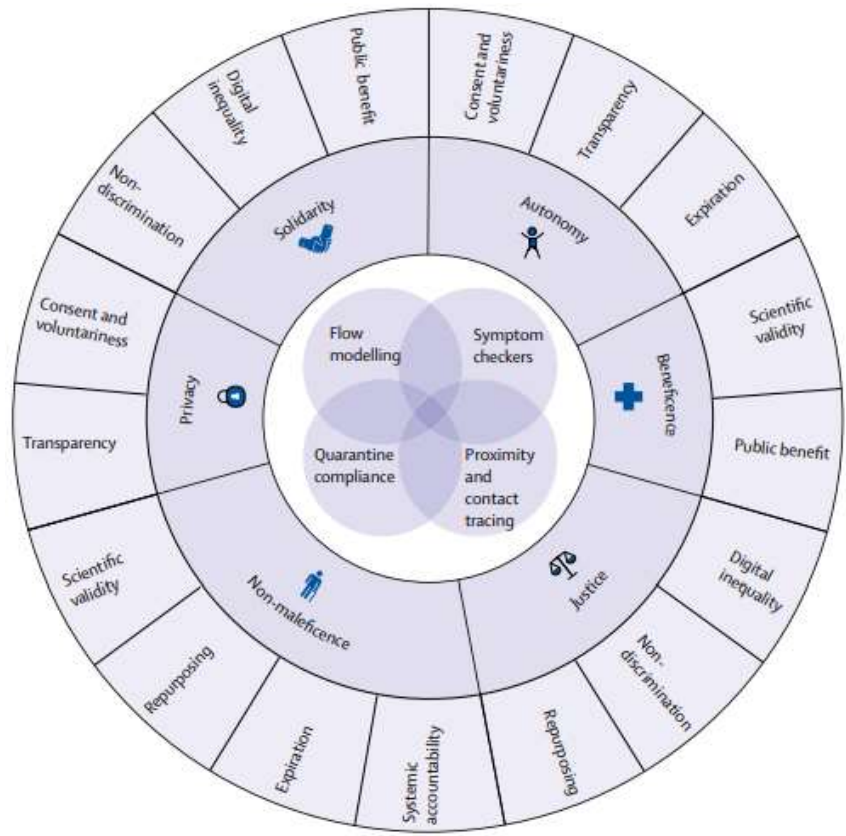

**Fig. 1.** The diagram mapping principles to digital health during COVID-19 technologies [26]

## 3    The world market for e-commerce

Internet connected 4 of 7 billion people. Global involvement is 52%, the majority of traffic falls on smartphones. The increase in the involvement of digital technologies has provoked an increase in investment in the Internet by 17%.

By the end of 2019, total sales of online stores of worldwide are predicted to reach the mark of $ 2 trillion. The increase compared to the year 2017, will be + 6%.

Almost half of all sales in e-commerce, following the results of 2017 will be in China (47%). In monetary terms, this is about $ 900 billion. Thus, the country will occupy the first place in the world in terms of sales in e-commerce, pushing at this point the United States.

The second largest regional market in e-commerce is North America. This year, sales in USA and other countries in the region are projected showing growth of + 15.6%. By 2020, the total sales of e-commerce in the world more than doubled, the COVID-19 pandemy is contributed to this process.

The chart of North America eHealth market from 2012 year to 2022 year, data in USD Million, is shown in Figure 5.

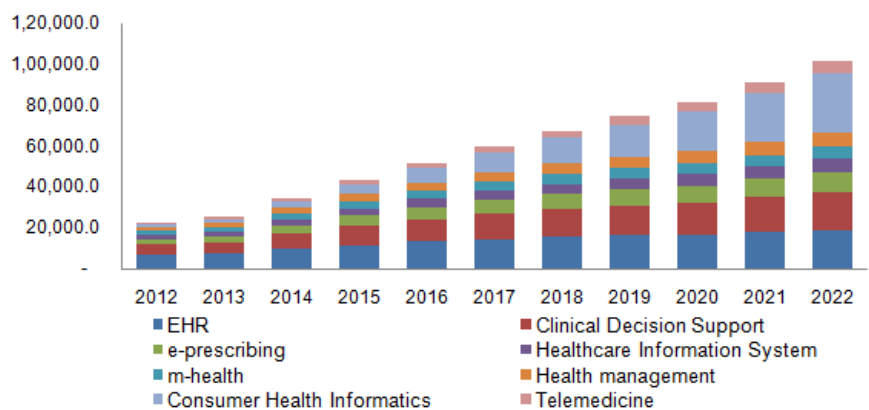

**Fig. 2.** The chart of North America eHealth market from 2012 year to 2022 year

In Europe, the share of e-commerce in GDP is quite high, and the highest rate was recorded in Britain 7.9%, followed by China 5.8% and France 3.59%. Moreover, the UK has the highest share of buyers who buy goods through electronic channels. Moreover, 81% of the country's population has access to the Internet, and 98% of them make online purchases. 81.5% of all online sales in Europe account for only three countries are the UK, Germany and France. And from year to year on this indicator, they show stable growth.

The center of e-commerce in Europe is the United Kingdom. The volume of online sales in the field of B2C is three times higher than in the second largest market on the continent – Germany. The leading platform here is Amazon, which accounts for 54% of all sales.

The main difficulties in the field of e-commerce, according to foreign sources, are the lack of security of data transmitted over the Internet, which is reduced to two threats – computer forgery and fraud and breach of privacy in the ability to clearly identify who, what, when and from whom he bought. Another fundamentally important problem is the quality of communications (development of the communication network, low channel capacity and low data rate).

### 3.1 Analysis of implamentation of e-commerce in the United Kingdom

The e-commerce market is $ 93.89 billion. 12.1% of purchases come from tablets, 16.5% from smartphones, and 71.4% from desktops. 33% of online sales take place after 6 p.m. E-commerce accounts for 30% of the country's economy. As for the share

of e-commerce in GDP, the the United Kingdohas the highest rate – 6.1%. Moreover, the the United Kingdom has the highest share of buyers who buy goods through electronic channels. 91.6% of the people have access to the Internet. The speed varies from 40 Mbit/s in cities up to 1 Mbit/s in remote locations. In this regard, the government has committed to provide 90% of households with high-speed broadband by 2020.

Today the most common payment methods in the the United Kingdom are credit and debit cards. E-wallets are not very well established in the country. With sales of $157 billion in 2014 (30% of all online sales in the region), the the United Kingdohas become the largest e-commerce market in Western Europe. 11% of all retail sales of the highest proportion in the whole of Western Europe are in online shopping accounts.

The chart of Internet sales as a proportion of products in the United Kingdom, 2016- 2019 [27] is shown in Fig. 4.

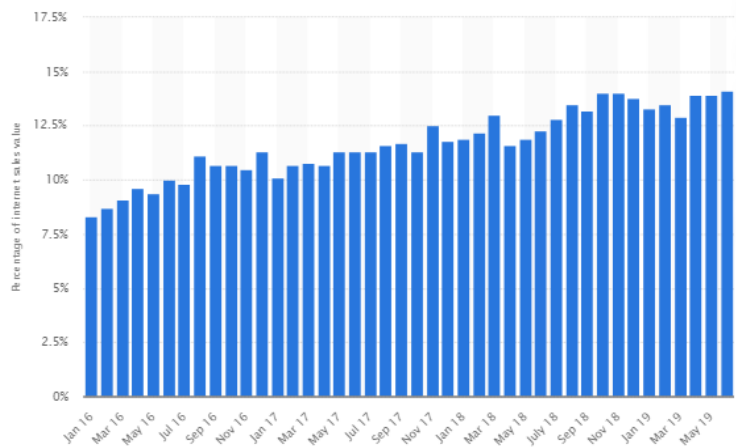

**Fig. 3.** The chart of Internet sales as a proportion of products stores in UK, 2016-2019

This statistic [27] shows internet sales as a proportion of products stores in UK, 2016-2019.

## 4     E-health

E-health (digital health) is an activity with the use of electronic information resources in the health care sector and ensuring prompt access of medical specialists and patients. The chart of digital health market [30] is shown in Figure 3.

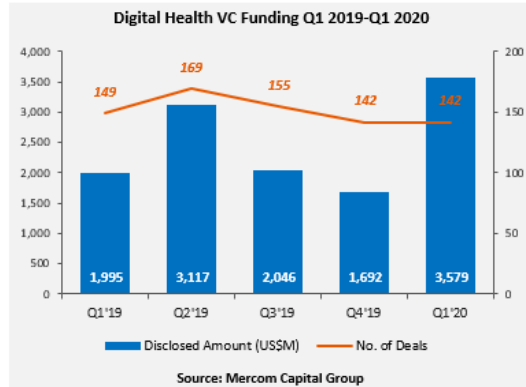

**Fig. 4.** The chart of digital health market (2019-2020)

According to compound annual growth rate Global Market Insights report the digital health market [30] is expected to reach $379B of 26% over 2017-2024.

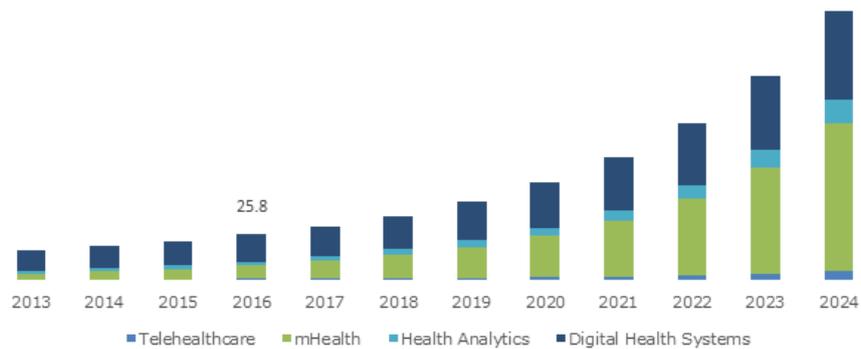

**Fig. 5.** The chart of Digital Health Market by 2024 year

The digital health landscape [32] is shown in Figure 1. According digital health landscape [32] the digital health is an covering field which includes sub-specialties such as:

― Blockchain
― Health IT
― Virtual and Augmented Reality
― Sensors and wearables
― Personal Genomics
― Telemedicine
― Big Data
― E-health
― Mobile Health
― Electronic Medical Record / Electronic Health Record
― Artificial Intelligence and Machine Learning

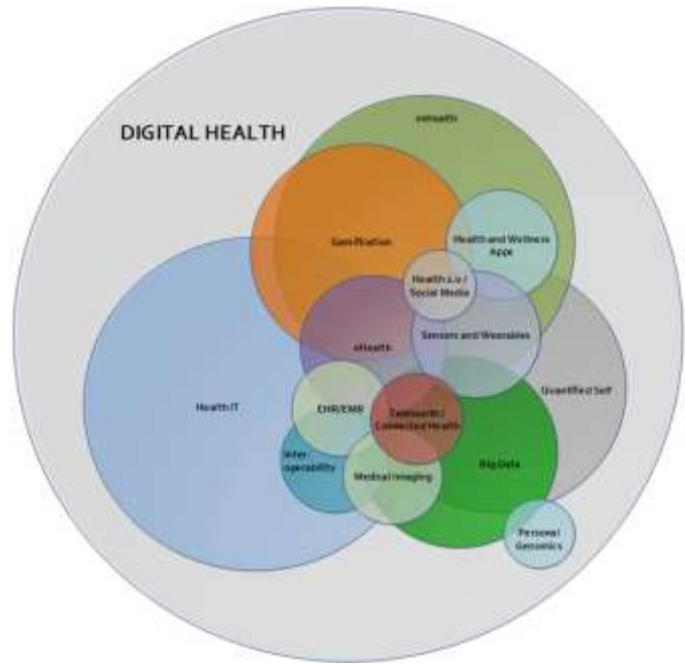

**Fig. 6.** The Digital Health Landscape

The popular sector is mobile health (mHealth). Mobile health is a using the health-related smartphone apps in the medicine and health sectors.

The chart of demand of mobile health in EU is presented in Figure 5.

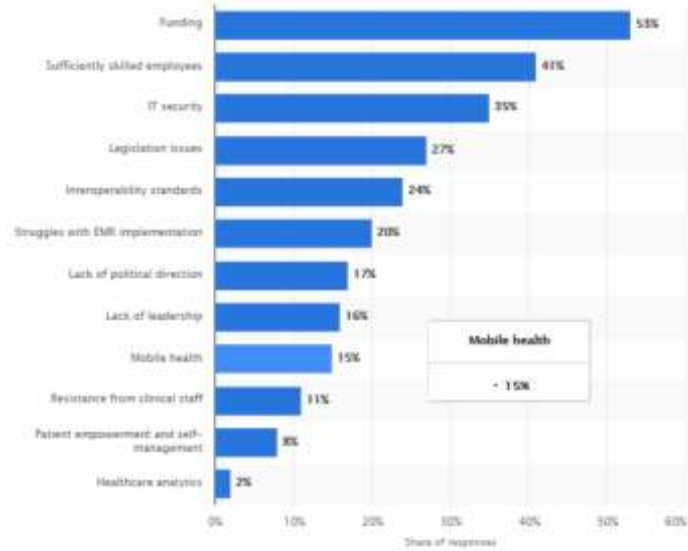

**Fig. 7.** The chart of demand of Mobile health in EU

The implementation of mHealth program in e-health has the following types: established, pilot or informal phase [13]. In chart mHealth program type by established, pilot or informal phase the condition of this type is presented.

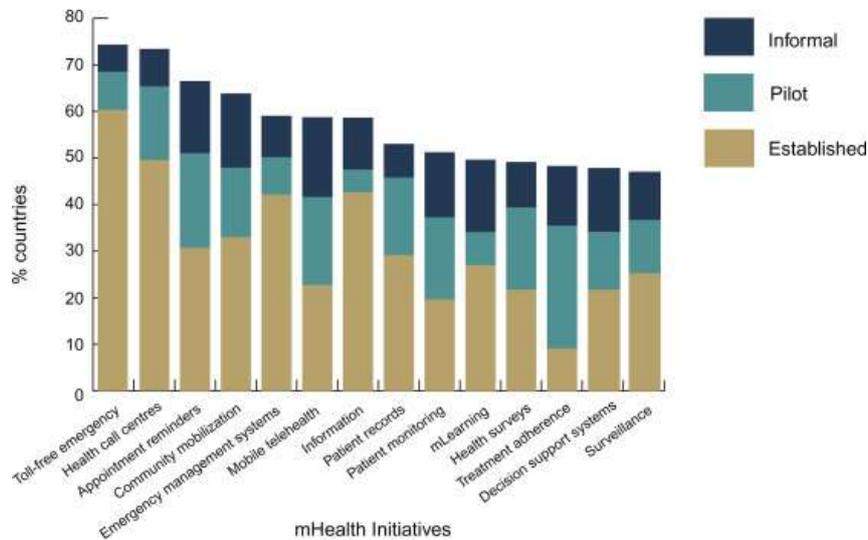

Fig. 8. Chart of mHealth program type by established, pilot or informal phase [13]

### 4.1 E-health in the United Kingdom

The aim of the Government of the United Kingdom is to introduce a wide-ranging program "System of electronic medical records in the United Kingdom by 2020".

The goal is that each patient's electronic record will contain data about their medical history of patient, treatment process, and lifestyle. Records should be available to all medical and social services. The participats can view and comment on a version of their medical record online. Patients can make appointments and repeat prescriptions online.

In the United Kingdom, e-health planning is implemented and implemented by the National Information Council, which develops imports and technologies for the national health care, i.e. social assistance; Information Center is a non-governmental organization that manages information, data and health information systems. These organizations develop national standards for electronic medicinal records, but native groups have opportunity to choose their own systems. NHS England funds several schemes, in order to encourage local groups to submit paper records to electronic databases.

The government supposes that initiatives will be dispersed, combined together and lead to complete coverage. By 2018, all clinical commission groups have already presented their plans, explaining introduction in e-health by 2020. The United Kingdom Health Authority has indicated future technology funds based on these plans. During the National IT Program (2002-16), the Ministry of Health collaborated with four vendors: CSC, BT, Accenture, and Fujitsu to provide electronic records. The NIB claims that the program was too centralized and insensitive to local circumstances.

Currently, the NHS stores information about patients on various media, ie both in paper and electronic form. Electronic records are kept by physicians (especially in radiology and pathology) as well as psychiatric hospitals. There is a great variation in the type of use of electronic recording systems, ranging from geographical regions to hospital units.

The advantages of e-medicine are that electronic records can provide primary and secondary data for patients, doctors. The Royal Academy of Medical Colleges has stated that the growing demands on health systems can only be met by the same development and use of electronic medical records. Involving patients in this system allows the patient to access data in different places, which is the key to ensuring the relationship between health care and the public.

## 5    Conclusions

The the experience of implementing e-commerce and e-health in word e-society is analysed. Condacted review study carries out the analysis of the experience of implementation of e-commerce and e-health. The investigation of e-commerce and e-health in the UK is proposed. The analysis of related works of e-commerce and e-health sectors is condacted. The world market for e-commerce is studied.

### References
1. Chaffey D., Hemphill T., Edmundson-Bird D. Digital business and e-commerce management. Pearson UK, 2019.
2. Laudon, Kenneth C., Carol Guercio Traver. E-commerce: business, technology, society. 2016.
3. Esser, Klaus, Judith Kurte. B2C e-commerce: impact on transport in urban areas. Recent Advances in City Logistics. The 4th International Conference on City LogisticsInstitute for City Logistics. 2006.
4. Khosrowpour, Mehdi, ed. The social and cognitive impacts of e-commerce on modern organizations. IGI Global, 2004.
5. Tachakra, Sapal, et al. Mobile e-health: the unwired evolution of telemedicine. Telemedicine Journal and E-health 9.3 (2003): 247-257.
6. Chan, Victor, Pradeep Ray, and Nandan Parameswaran. Mobile e-Health monitoring: an agent-based approach. IET communications 2.2 (2008): 223-230.
7. Archer, Norm. Mobile e-health: Making the case. Mobile government: An emerging direction in E-government. IGI Global, 2007. 155-170.
8. Pappas, Christodulos, et al. A mobile e-health system based on workflow automation tools. Proceedings of Symposium on Computer-Based Medical Systems (CBMS 2002). 2002.
9. Peddi, Sri Vijay Bharat, et al. An intelligent cloud-based data processing broker for mobile e-health multimedia applications. Future Generation Computer Systems 66: 71-86 (2017).
10. Perera, Chandrashan. The evolution of E-Health–mobile technology and mHealth. Journal of Mobile Technology in Medicine 1.1 (2012): 1-2.

11. Mackert, Michael, Brad Love, Pamela Whitten. Patient education on mobile devices: an e-health intervention for low health literate audiences. Journal of Information Science 35.1 (2009): 82-93.
12. Aramudhan, M., Mohan, K.. New secure communication protocols for mobile e-health system. International Conference on Networked Digital Technologies. Springer, Berlin, Heidelberg, 2010.
13. Pinciroli, Francesco, et al. Telemedicine and e-health. IEEE pulse 2.3 (2011): 62-70.
14. Maheu, Marlene, Pamela Whitten, Ace Allen. E-Health, Telehealth, and Telemedicine: a guide to startup and success. John Wiley & Sons, 2002.
15. Riva, Giuseppe. From telehealth to e-health: Internet and distributed virtual reality in health care. CyberPsychology & Behavior 3.6 (2000): 989-998.
16. Son, Youn-Jung, et al. Visualization of e-Health research topics and current trends using social network analysis. Telemedicine and e-Health 21.5 (2015): 436-442.
17. Wootton, Richard. Telemedicine. Bmj 323.7312 (2001): 557-560.
18. Perednia, Douglas A., Ace Allen. Telemedicine technology and clinical applications. Jama 273.6 (1995): 483-488.
19. Hjelm, N. Benefits and drawbacks of telemedicine. Journal of telemedicine and telecare 11.2 (2005): 60-70.
20. Hamed A. Essentials of telemedicine and telecare. Chichester: Wiley, 2002.
21. Ekeland A., et al. Effectiveness of telemedicine: a systematic review of reviews. International journal of medical informatics 79.11 (2010): 736-771.
22. Tarassenko, Lionel, et al. Telemedicine system. U.S. Patent Application No. 10/528,365.
23. Craig J., Petterson V. Introduction to the practice of telemedicine. Journal of telemedicine and telecare 11.1 (2005): 3-9.
24. Bashshur, Rashid L. On the definition and evaluation of telemedicine. Telemedicine Journal 1.1 (1995): 19-30.
25. Pattichis, Constantinos S., et al. Wireless telemedicine systems: an overview. IEEE Antennas and Propagation Magazine 44.2 (2002): 143-153.
26. Gasser, U., et al. Digital tools against COVID-19: taxonomy, ethical challenges, and navigation aid (2020). https://www.thelancet.com/action/showPdf?pii=S2589-7500%2820%2930137-0
27. Internet sales as a proportion of all retailing in household goods stores in the United Kingdom. https://www.statista.com/statistics/982882/ internet-sales-as-proportion-of-all-retailing-household-goods-stores-uk/
28. Digital Health Report (2020) https://mercomcapital.com/product/q1-2020-digital-health-healthcare-it-funding-ma-report/
29. World Health Statistics 2020. A visual summary. https://www.who.int/data/gho/whs-2020-visual-summary
30. Digital Health Market. https://hitconsultant.net/2018/06/18/digital-health-market-2024/#.Xwh1jygzbIU
31. eHealth Market Analysis. https://www.grandviewresearch.com/industry-analysis/e-health-market
32. The Digital Health Landscape. https://innovatemedtec.com/digital-health
    Hofmaier S., Huang X., Matricardi P. M., Telemedicine and Mobile Health Technology in the Diagnosis, Monitoring and Treatment of Respiratory Allergies. Implementing Precision Medicine in Best Practices of Chronic Airway Diseases. (2019). 117-124.